\documentclass[pss,fleqn]{w-art}
\usepackage{times}
\usepackage{w-thm}
\usepackage[]{graphicx}
\begin{document}
\DOIsuffix{theDOIsuffix}
\Volume{XX}
\Issue{1}
\Month{01}
\Year{2003}
\pagespan{1}{}
\Receiveddate{\sf } 
\Reviseddate{\sf  } 
\Accepteddate{\sf  } 
\Dateposted{\sf  }
\keywords{stripe phase, Hubbard model, doped cuprates.}
\subjclass[pacs]{71.10.Fd, 71.27.+a, 74.72.-h}



\title[Half-filled stripes in the $t$-$t'$-$U$  Hubbard model]
{Half-filled stripes in the $t$-$t'$-$U$  Hubbard model}


\author[M. Raczkowski] {Marcin Raczkowski\footnote{Corresponding
     author: e-mail: {\sf M.Raczkowski@if.uj.edu.pl},
     Phone: +48\,12\,663\,5628,
     Fax: +48\,12\,633\,4079}\inst{1,2}}
\address[\inst{1}] {Marian Smoluchowski Institute of Physics, Jagellonian
 University, Reymonta 4, 30059 Krak\'ow, Poland}
\address[\inst{2}] {Laboratoire CRISMAT, UMR CNRS-ENSICAEN (ISMRA) 6508, 
14050 Caen, France}

\author[A.~M. Ole\'s]{Andrzej M. Ole\'s\inst{1}}

\author[R. Fr\'esard]{Raymond Fr\'esard\inst{2}}

\begin{abstract}
Using a self-consistent Hartree-Fock approximation we investigate
the relative stability of various stripe phases in the extended
$t$-$t'$-$U$ Hubbard model. One finds that a negative ratio of next- to 
nearest-neighbor hopping $t'/t<0$ expells holes from antiferromagnetic 
domains and reinforces the stripe order. Therefore the half-filled 
stripes not only accommodate holes but also redistribute them so that 
the kinetic energy is gained, and these stripes take over in the regime 
of $t'/t\simeq -0.3$ appropriate for YBa$_2$Cu$_3$O$_{6+\delta}$.

\end{abstract}
\maketitle                   

Two-dimensional (2D) systems with strong electron correlations received 
a great deal of attention over the past twenty years. Such systems 
exhibit a subtle competition between rich variety of phases: 
antiferromagnetic (AF), ferromagnetic (FM), charge density wave (CDW), 
spin density wave (SDW), and superconducting phase. In particular, strong 
interplay of different order parameters in doped transition metal oxides 
leads to the experimentally observed phase separation involving holes 
ordered into domain walls (DWs) separated by AF domains \cite{Kiv03}. 
On the theoretical side, stripe phases have been predicted on the basis 
of Hartree-Fock (HF) calculations of doped AF clusters \cite{Zaa89}. 
One important feature of these results is that they indicate formation of 
the so-called {\it filled stripes\/} with density of one doped hole per 
one atom in a DW. However, neutron  experiments on the Nd-codoped
La$_{2-x}$Sr$_x$CuO$_4$ (Nd-LSCO) cuprates reveal that the observed 
stripes are filled by one hole per two DWs, corresponding to  
{\it half-filled stripes\/} \cite{Tra95Nd}. Unfortunately, the latter are 
only locally stable within the HF approximation \cite{Zaa96}. Even though 
some additional Hamiltonian terms like the nearest-neighbor Coulomb 
repulsion $V$ may slightly enhance their stability, filled stripes remain 
always better solutions \cite{Zaa96}. Nonetheless, we will show below 
that under some circumstances, half-filled vertical site-centered (HVSC) 
stripes may become energetically favored over the filled either diagonal 
site-centered (DSC) or vertical site-centered (VSC) ones even in the HF 
approximation.

\begin{figure}[t!]
\begin{center}
\includegraphics[width=0.7\textwidth]{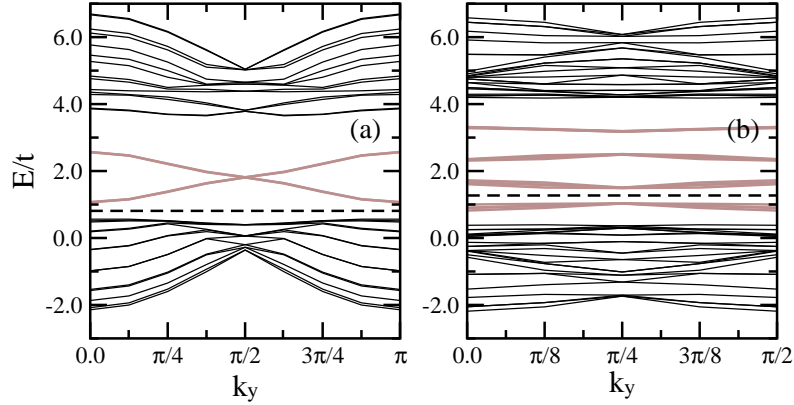}
\end{center}
\caption
{
Band structure as a function of momentum $k_y$, for
a unit cell of width $N_x=16$ in the Hubbard model with $U/t=5$ for:
(a) VSC stripe, doubled Brillouin zone; 
(b) HVSC stripe, quadrupled Brillouin zone.  
Black (gray) line corresponds to the bulk (mid-gap) bands, 
respectively, whereas the dashed line indicates the Fermi level. 
}
\label{fig1}
\end{figure}

The starting point for the analysis of stripe structures is the effective 2D
single-band Hubbard model,
\begin{equation}
H=-\sum_{ij\sigma}t^{}_{ij}c^{\dag}_{i\sigma}c^{}_{j\sigma} +
  U\sum_{i}n^{}_{i\uparrow}n^{}_{i\downarrow},
\label{eq:Hubb}
\end{equation}
where the hopping $t_{ij}$ is $t$ on the bonds connecting nearest-neighbor
sites and $t'$ for next-nearest-neighbor sites, while $U$ is the on-site 
Coulomb interaction. 
The model is solved self-consistently within the HF method on 
a $16\times 16$ cluster with periodic boundary conditions. For a 
representative doping level $x=1/8$ one obtains stable filled stripe 
structures with AF domains of width seven atoms. Their stability can be 
best understood from the band structure shown in Fig.~\ref{fig1}.  
In general, each filled DW induces the formation of two unoccupied bands, 
i.e., one for the $\uparrow$-spin and another one for the 
$\downarrow$-spin, lying within the Mott-Hubbard gap, as illustrated in 
Fig.~\ref{fig1}(a). Consequently, their special stability rests on a gap
that opens in the symmetry broken state between the highest occupied 
state of the lower Hubbard band and the bottom of the mid-gap band. One 
may ask what happens now if one decreases the doping level so as to get 
half-filled stripes without a period quadrupling? Certainly, the twofold 
degenerate mid-gap states become quarter filled, as there is only half a 
hole per one DW. Hence, any stringent reason for such a symmetry breaking 
with the Fermi level in the middle of the lower mid-gap band is absent.
Therefore, one needs to lower the symmetry by a period quadrupling so 
that the period on the stripe becomes four times the lattice constant 
\cite{Zaa96}. This might be accomplished by introducing a SDW modulation 
along the DWs which leads to the band structure shown in Fig.~\ref{fig1}(b), 
with a gap that opens up exactly at the Fermi energy. However, in spite of
better optimizing the potential energy $E_U$, the HVSC stripe phase 
represents only a local minimum of energy, being less stable than both the 
filled VSC and DSC ones. Thus, guided by the observation that finite 
negative $t'$ expels holes from the AF domains and reinforces the stripe 
order \cite{Rac03}, we investigate whether this mechanism suffices to 
stabilize the former structure in the $t$-$t'$-$U$ model. 

\begin{table}[b!]
\caption {Energies per site: ground-state energy $E_{\rm tot}$, kinetic 
energy contributions for the bonds along (10) $E_t^{x}$, (01) $E_t^{y}$, 
(11) $E_{t'}^{x-y}$ and ($1\bar{1}$) $E_{t'}^{x+y}$ directions, 
and potential energy $E_U$ of the HVSC, VSC and DSC stripe phases in the 
$t$-$t'$-$U$ model with $U/t=5$ and $x=1/8$. VSC stripe is unstable at 
$t'/t=-0.3$.}
\label{tab1}
\begin{tabular}{rrccrrcc}
\hline
\multicolumn{1}{c} {}             &\multicolumn{1}{c}  {$t'/t$}        & 
\multicolumn{1}{c} {$E_t^x/t$}    &\multicolumn{1}{c}  {$E_t^y/t$}     & 
\multicolumn{1}{c} {$E_{t'}^{x-y}/t$}&\multicolumn{1}{c} {$E_{t'}^{x+y}/t$}  & 
\multicolumn{1}{c} {$E_U/t$}      &\multicolumn{1}{c} {$E_{\textrm{tot}}/t$}\\ 
\hline
{\bf HVSC}  &  0.0       & $-$0.5846   & $-$0.6585  & 0.0000 
            &  0.0000    &    0.4627   & $-$0.7804   \\
{\bf VSC}   &  0.0       & $-$0.6753   & $-$0.6147  & 0.0000 
            &  0.0000    &    0.4900   & $-$0.8000   \\ 
{\bf DSC}   &  0.0       & $-$0.6368   & $-$0.6368  &  0.0000  
            &  0.0000    &    0.4696   & $-$0.8040   \\
\hline
{\bf DSC}   &  $-$0.3    & $-$0.6152   & $-$0.6152  &  0.0000  
            &  0.0257    &    0.4379   & $-$0.7668   \\    
{\bf HVSC}  &  $-$0.3    & $-$0.5786   & $-$0.6198  & $-$0.0080  
            &  $-$0.0080 &    0.4322   & $-$0.7678   \\   
\hline
\end{tabular}
\end{table}

\begin{figure}[t!]
\begin{minipage}{.57\textwidth}
\includegraphics[width=.6\textwidth]{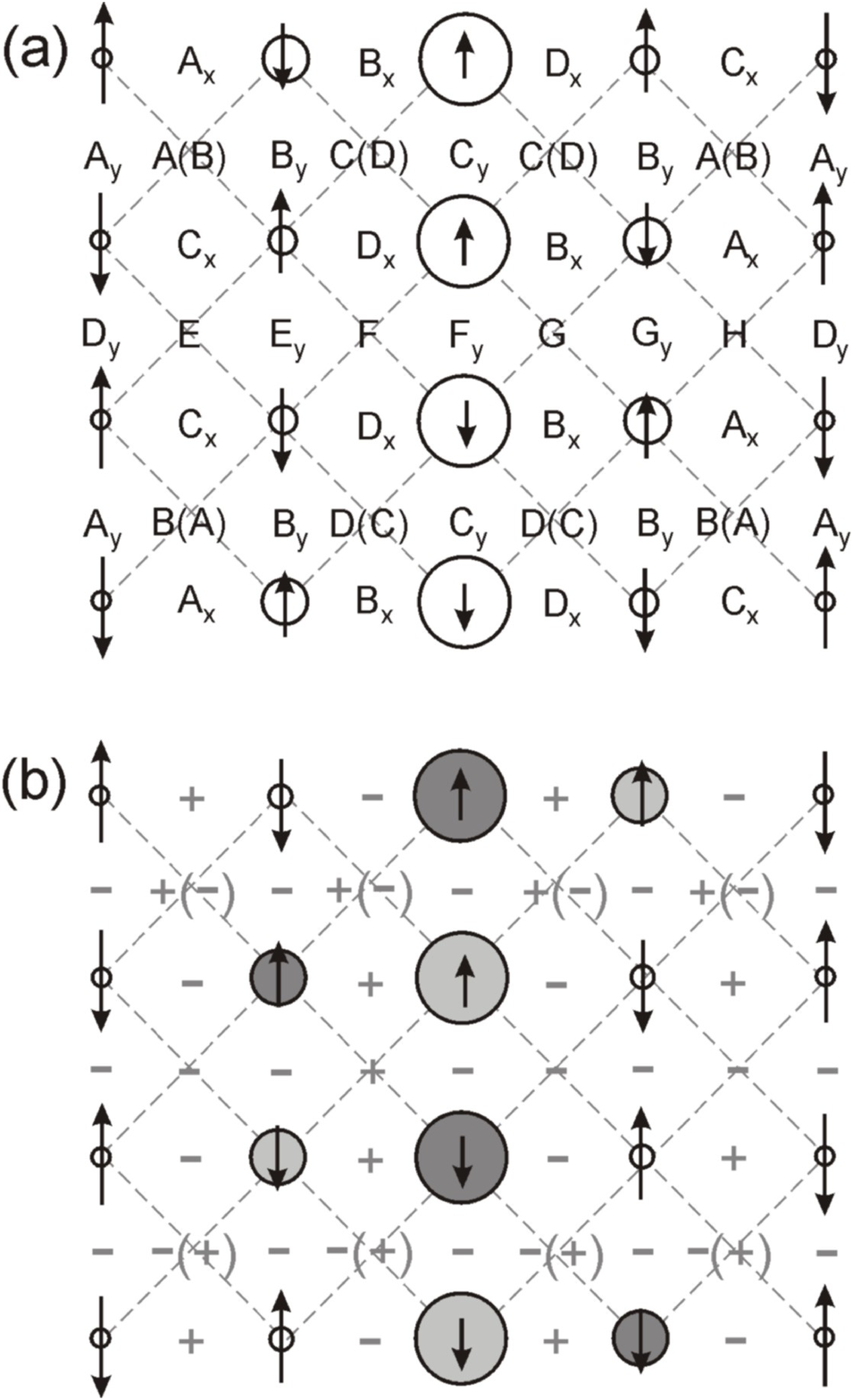}
\caption
{
Rearrangement of the HVSC stripe order with increasing 
next-nearest-neighbor hopping: 
(a) $t'/t=0$; 
(b) $|t'|/t=0.45$. Circle sizes are proportional to doped hole 
density. Capital letters label bond kinetic energies along the 
$(10)$ (A$_x,\dotsc,$D$_x$), $(01)$ (A$_y,\dotsc,$G$_y$), 
$(11)$ (A,$\dotsc$,H), and $(1{\bar 1})$ (in parenthesis) directions, 
listed in Table~\ref{tab2}, while $+$ ($-$) correspond to the bond 
kinetic energy gain (loss) with respect to the $t'=0$ case.
Shadow circles indicate two crossing zigzag lines, see text 
($U/t$=5, $x=1/8$). 
}
\label{fig2}
\end{minipage}
\hfil
\begin{minipage}{.35\textwidth}
\setfloattype{table}
\caption
{
Bond kinetic energies along the $(10)$ (A$_x,\dotsc,$D$_x$), $(01)$ 
(A$_y,\dotsc,$G$_y$), $(11)$ and $(1{\bar 1}$) (A,$\dotsc$,H) 
directions around the HVSC stripe shown in Fig.~\ref{fig2}. 
}
\label{tab2}
\begin{tabular}{crr}
\hline
\multicolumn{1}{c|}{}& \multicolumn{2}{c}{$t'/t$ }     \\
\cline{2-3}
\multicolumn{1}{c|}{bond}     & \multicolumn{1}{c}{0 }   &
\multicolumn{1}{c}{$-0.45$}   \\
\hline
A$_x$ &  $-$0.6081 & $-$0.6141 \\
B$_x$ &  $-$0.6652 & $-$0.5482 \\
C$_x$ &  $-$0.6277 & $-$0.6257 \\
D$_x$ &  $-$0.4377 & $-$0.4885 \\
\hline
A$_y$ &  $-$0.6057 & $-$0.5822  \\
B$_y$ &  $-$0.6003 & $-$0.5281  \\
C$_y$ &  $-$0.8505 & $-$0.7234  \\
D$_y$ &  $-$0.5759 & $-$0.5623  \\
E$_y$ &  $-$0.6515 & $-$0.5299  \\
F$_y$ &  $-$0.7206 & $-$0.6364  \\
G$_y$ &  $-$0.6635 & $-$0.6046  \\
\hline
A &  0.0000 & $-$0.0053    \\
B &  0.0000 &    0.0262    \\
C &  0.0000 & $-$0.1300    \\
D &  0.0000 &    0.0128    \\
E &  0.0000 &    0.0033    \\
F &  0.0000 & $-$0.0367    \\
G &  0.0000 &    0.0164    \\
H &  0.0000 &    0.0020    \\
\hline
\end{tabular}
\end{minipage}
\end{figure}

In order to establish the role of $t'$, we begin with energies of 
different phases at $t'=0$ and $t'=-0.3t$ (Table \ref{tab1}). The main 
effect is the energy gain of HVSC over DSC phase along (11) and 
$(1\bar{1})$ bonds with increasing $|t'|$. To understand it better we 
discuss Fig.~\ref{fig2} showing how the charge and spin configuration 
around the HVSC stripe is altered in the presence of increasing $|t'|$. 
In Fig. \ref{fig2}(a) we label all inequivalent bond kinetic energies 
along the $(10)$ (A$_x,\dotsc,$D$_x$), $(01)$ (A$_y,\dotsc,$G$_y$), 
$(11)$ (A,$\dotsc$,H), and $(1{\bar 1})$ (in parenthesis) directions. 
The most striking feature of the HVSC stripe phase is that the main 
kinetic energy gain is released not by transverse charge fluctuations, 
which is the case for the filled stripes, but by on-wall hopping 
processes on the bonds $C_y$ ($F_y$) connecting sites with the FM (AF) 
coupling, respectively (see Table~\ref{tab2}). Another interesting 
property is that the system tries to regain some of the kinetic energy 
in the transverse direction by developing a CDW on nearest-neighbor 
sites to the DW itself. Indeed, although the charge distribution along 
the stripe is uniform, the on-wall SDW causes the CDW on both sides of 
the DW. This clearly promotes the hopping between the AF sites (bonds 
$B_x$) over the FM ones (bonds $D_x$). Furthermore, the CDW is strongly 
influenced by $t'$, being first almost entirely quenched when 
$t'\simeq -0.15$ and then it is gradually restored [\textit{cf}. Fig.
\ref{fig2}(b)]. 

These trends are fully consistent with the dynamical mean field theory 
studies of the effect of a single kink along the HVSC stripe \cite{Fle01}. 
It has been established that the energy cost of forming a kink increases 
slightly with increasing $|t'|$ up to $|t'|/t<0.1$, whereas further 
increase leads to the opposite effect, and finally a wall with a kink 
becomes favored at $|t'|/t\simeq 0.3$. Note, however, that although the 
overall shape of the brought back CDW is the same as of the initial one, 
the physical situation is fundamentally different -- in the $t'=0$ limit, 
it is energetically advantageous for the system to equalize hole density 
between sites connected by the bonds $B_x$ (AF coupling), whereas in 
the large $|t'|$ limit, it tries instead to equalize hole density between 
sites connected by the bonds $D_x$ (FM coupling). Accordingly, such a  
charge redistribution results in two crossing zigzag lines, 
which facilitate hole propagation. Indeed, each path consists of 
a -$C$-$F$- bond pattern along the (11) direction and analogous sequence 
in the $(1\bar{1})$ direction. Even though a moving hole gains kinetic 
energy on both type of bonds, a more significant gain is achieved on the 
FM bond $C$, being the driving force in the formation of zigzag pattern. 
It is also supported by a systematically growing kinetic energy gain 
with increasing $|t'|$ on the FM bond $D_x$ connecting the zigzag paths 
(\textit{cf}. Table~\ref{tab2}).

\begin{figure}[t!]
\includegraphics[width=0.9\textwidth]{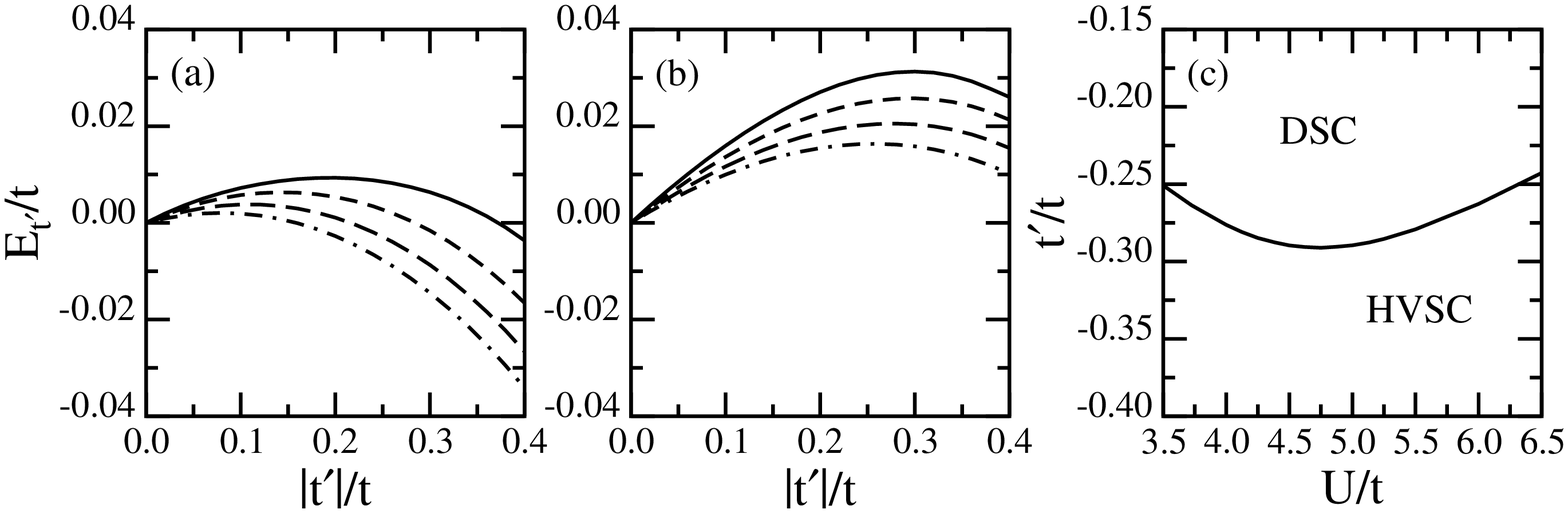}
\caption
{
Average kinetic energy $E_{t'}$ per diagonal bond as a function of 
increasing next-nearest-neighbor hopping $|t'|$ for: 
(a) half-filled vertical site-centered (HVSC) stripe phase; 
(b) filled diagonal site-centered (DSC) stripe phase, as obtained for
$U/t=4$, 5, 6, 7 (solid, dashed, long-dashed, dot-dashed line).
Panel (c) shows the phase boundary between the DSC stripe phase and the 
HVSC one as obtained for doping $x=1/8$. 
}
\label{fig3}
\end{figure}

The effect of increasing $|t'|$ on the kinetic energy per diagonal bond 
$E_t'$ of the HVSC stripe phase illustrates Fig.~\ref{fig3}(a). One 
observes that in the small $|t'|$ regime next-nearest-neighbor hopping 
processes cost energy and therefore they are suppressed by quenching 
the CDW order on sites next to the DW. However, towards larger $|t'|$, 
the kinetic energy associated with next-nearest-neighbor hopping 
becomes negative despite the negative sign of $t'$. 
Consequently, the system develops a new CDW so as 
to optimize the $E_{t'}$ gain. In contrast, $E_{t'}$ of the filled DSC 
stripe phase remains positive even in the large $|t'|=0.4t$ regime, as 
shown in Fig.~\ref{fig3}(b). Moreover, increasing Coulomb repulsion $U$ 
yields in this case only a small suppression of $E_{t'}$, whereas the 
same increase of $U$ is reflected in much larger $E_{t'}$ gain of the 
HVSC stripe phases. This, together with a better on-site energy $E_U$, 
explains a broader region of stability of the HVSC stripe phases in the 
strongly correlated regime, as illustrated in Fig.~\ref{fig3}(c). 
We interprete their enhanced stability in the small $U$ regime 
as following from the melting of both stripe phases. 
Indeed, increased mobility of the holes released 
from the DSC stripes results in a faster, as compared to the HVSC case, 
enhancement of the $E_{t'}$ energy cost upon increasing $|t'|$.

In conclusion, we have found that the next-nearest-neighbor hopping $t'$ plays 
an important role in affecting the relative stability between filled 
and half-filled stripe phases. However, the established value 
$|t'|/t\simeq 0.3$ opening a window for the stability of the HVSC stripe 
phase, is excessively larger than $t'/t\simeq -0.1$ of LSCO compounds 
\cite{Pav01}. Interestingly, $t'/t\simeq -0.3$ corresponds to the value 
appropriate for YBa$_2$Cu$_3$O$_{6+\delta}$ (YBCO) and indeed a 
half-filled vertical stripe phase has been reported in this compound for 
$\delta=0.35$ \cite{Moo02}. Therefore, our result calls for a further 
experimental characterization of stripes in YBCO. Finally, 
it may be expected that a better treatment of electron 
correlations would stabilize further the half-filled stripe phases 
\cite{Sei04} for realistic values of $t'$.

\begin{acknowledgement}
We thank P. Dai for valuable discussions. M. Raczkowski was supported 
by a Marie Curie Fellowship of the European Community program under 
number HPMT2000-141. This work was also supported by the Polish 
Ministry of Scientific Research and Information Technology, 
Project No. 1~P03B~068~26.
\end{acknowledgement}



\begin{thebibliography}{10}

\bibitem{Kiv03} S.~A. Kivelson {\it et al.},    
                   Rev. Mod. Phys. {\bf 75}, 1201 (2003).

\bibitem{Zaa89} J.~Zaanen and  O.~Gunnarsson,
                   Phys. Rev. B {\bf 40}, 7391 (1989); 
		D. Poilblanc and T. M. Rice, 
		   {\it ibid.} {\bf 39}, 9749 (1989). 

\bibitem{Tra95Nd} J.~M.~Tranquada, B.~J.~Sternlieb, J.~D.~Axe,
                  Y.~Nakamura, and  S.~Uchida,
                   Nature {\bf 375}, 561 (1995).

\bibitem{Zaa96} J. Zaanen and A.~M. Ole\'s,
                   Ann. Phys. (Leipzig) {\bf 5}, 224 (1996).

\bibitem{Rac03} M. Raczkowski, B. Normand, and A.~M. Ole\'s,
                   Phys. Stat. Sol. (b) \textbf{236}, 376 (2003).

\bibitem{Fle01} M.~Fleck, A.~I.~Lichtenstein, and A.~M.~Ole\'s, 
                   Phys. Rev. B {\bf 64}, 134528 (2001).

\bibitem{Pav01} E. Pavarini, I. Dasgupta, T. Saha-Dasgupta, O. Jepsen, 
                   and O. K. Andersen, 
                   Phys. Rev. Lett. {\bf 87}, 047003 (2001).

\bibitem{Moo02} H.~A. Mook, P. Dai, and F. Do\v{g}an,
                   Phys. Rev. Lett. {\bf 88}, 097004 (2002).

\bibitem{Sei04} G. Seibold and J. Lorenzana,
                   Phys. Rev. B {\bf 69}, 134513 (2004).

\end{thebibliography}
\end{document}